# Gravitational field of a higher dimensional global monopole in Lyra geometry

F. Rahaman, K. Maity, P. Ghosh and K. Gayen

Dept. of Mathematics, Jadavpur University, Kolkata-700032, India

E-Mail: farook_rahaman@yahoo.com

**Abstract:**
We present a five dimensional global monopole within the framework of Lyra's geometry. Also the gravitational field of the monopole solution has been considered.

## Introduction:

In quantum field theory, when a symmetry has been broken during the phase transition, then several topological defects will arise [1]. Monopoles are point like topological defects that may arise during phase transitions in the early Universe. In particular, $\pi_2(M) \neq I$, ( M is the vacuum manifold ) i.e. M contains surfaces which can not be continuously shrunk to a point, then monopoles are formed [2]. Global monopoles are formed from breaking of a global $SO(3)$ symmetry. Such monopoles have goldstone fields with energy density decreasing with the distance as inverse square law. Hence the total energy in the Goldstone field surrounding global monopoles suggests that they can produce strong gravitational fields [3].

At first, Barriola and Vilenkin (BV) [4] have obtained an approximate solution of the Einstein equations for the static metric out side a global monopole. They considered the monopole as associated with a triplet of scalar fields:

$$\varphi^a = \eta\, f(r)\, [x^a / r] \qquad \{ a = 1,2,3 \} \qquad \ldots\ldots(1)$$

$$(\text{with } x^a x^a = r^2)$$

With the static spherically symmetric line element,

$$ds^2 = e^\gamma dt^2 - e^\mu dr^2 - r^2 d\Omega_2^2 \qquad \ldots(2)$$

$f(r)$ is to obey the equation [Eular-Lagrange equation]

$$e^{-\mu} f^{11} + e^{-\mu} [(2/r) + \tfrac{1}{2}\gamma^1 - \tfrac{1}{2}\mu^1] f^1 - (2f/r^2) - \lambda\eta^2 f(f^2 - 1) = 0 \qquad \ldots(3)$$

( where '1' denotes differentiation w.r.t. r )

BV then assumed f = 1 out side the core.
Clearly this is not a solution of (3) – as an approximate solution, it requires the neglect of all powers of $(1/r^2)$ in the flat space.
To get a more correct picture about monopole, Harrari and Luosto (HL) [5] has considered

$$f(r) = 1 - (1/x^2) - [\{(3/2) - \Delta\}/x^4] + 0(x^{-6}) \quad \ldots(4)$$

where $x = \sqrt{\lambda}\eta r$, $\Delta = 8\pi G \eta^2$

It satisfies (3) up to $0(r^{-2})$ in the flat space.

The unification of gravitational force with the other forces in nature is not possible in the usual 4 – dimensional space-time. So higher dimensional theory might be useful at the very early stages of the evolution of the Universe. Solutions of Einstein field equations in higher dimensional space times are believed to be of physically relevance possibly at the extremely early times before the Universe underwent compactification transitions.

In last few decades there has been considerable interest in alternative theory of gravitation. The most important among them being scalar tensor theories proposed by Lyra [6] and Brans-Dicke [6]. Lyra proposed a modification of Riemannian geometry by introducing a gauge function into the structure less manifold that bears a close resemblances to Weyl's geometry. In consecutive investigations Sen [7] and Sen and Dunn [7] proposed a new scalar tensor theory of gravitation and constructed an analog of the Einstein field equation based on Lyra's geometry which in normal gauge may be written as

$$R_{ik} - \tfrac{1}{2} g_{ik} R + (3/2) \phi_i \phi_k - \tfrac{3}{4} g_{ik} \phi_m \phi^m = -8\pi T_{ik} \quad \ldots(5)$$

where $\phi_i$ is the displacement vector and other symbols have their usual meaning as in Riemannian geometry.
Halford [8] has pointed out that the constant displacement field $\phi_i$ in Lyra's geometry play the role of cosmological constant $\Lambda$ in the normal general relativistic treatment. According to Halford the present theory predicts the same effects within observational limits, as far as the classical solar system tests are concerned, as well as tests based on the linearised form of field equations.
Subsequent investigations were done by several authors in scalar tensor theory and cosmology within the framework of Lyra geometry [8].
In recent, Farook [9] and Farook et al [10] have studied some topological defects within the framework of Lyra geometry. In recent past Banerji et al [11] have extended the work of BV in to the five dimensional space-time.
In this paper, we would like to discuss higher dimensional global monopole with constant displacement vector based on Lyra geometry in normal gauge i.e. displacement vector $\phi_i = (\beta_0, 0, 0, 0, 0)$ where $\beta_0$ is a constant and look forward whether the monopole shows any significant properties due to introduction of the gauge field in the Riemannian geometry.

## 2. Basic equations:

The most general five dimensional static metric admitting spherically symmetry in the usual three dimensional space is taken to be

$$ds^2 = e^\gamma dt^2 - e^\beta dr^2 - r^2 d\Omega_2^2 - e^\mu d\psi^2 \qquad ....(6)$$

Here $\gamma, \beta, \mu$ are functions of r alone and $\psi$ is the fifth co ordinate .

The energy momentum tensors can be written via

$$T^{ab} = 2 (\partial L / \partial g_{ab}) - L g^{ab} \qquad ......(7)$$

with the Lagrangian $L = \frac{1}{2} \partial_\mu \varphi^a \partial^\mu \varphi^a - \frac{1}{4} \lambda (\varphi^a \varphi^a - \eta^2)^2$ as follows

$$T_t^t = \{\eta^2 (f^1)^2 / 2 e^\beta\} + (\eta^2 f^2 / r^2) + \frac{1}{4} \lambda (\eta^2 f^2 - \eta^2)^2 \qquad ........(8)$$

$$T_r^r = - \{\eta^2 (f^1)^2 / 2 e^\beta\} + (\eta^2 f^2 / r^2) + \frac{1}{4} \lambda (\eta^2 f^2 - \eta^2)^2 \qquad ........(9)$$

$$T_\theta^\theta = T_\varphi^\varphi = \{\eta^2 (f^1)^2 / 2 e^\beta\} + \frac{1}{4} \lambda (\eta^2 f^2 - \eta^2)^2 \qquad ........(10)$$

$$T_\psi^\psi = \{\eta^2 (f^1)^2 / 2 e^\beta\} + (\eta^2 f^2 / r^2) + \frac{1}{4} \lambda (\eta^2 f^2 - \eta^2)^2 \qquad .....(11)$$

It can be shown that in flat space the monopole core has a size $\delta \sim \sqrt{\lambda} \eta^{-1}$ and mass, $M_{core} \sim \lambda^{-1/2} \eta$. Thus if $\eta << m_p$ where $m_p$ is the plank mass, it is evident that we can still apply the flat space approximation of $\delta$ and $M_{core}$ [11].
This follows from the fact that in this case the gravity would not much influence on monopole structure.

Out side the core, however, one can approximate [5]

$f(r) = 1 - (1 / x^2) - [\{(3/2) - \Delta\} / x^4] + 0 (x^{-6})$, and then energy momentum tensors simplify to

$$T_t^t = T_r^r = T_\psi^\psi = (\eta^2 / r^2) - (1 / \lambda r^4) ; \quad T_\theta^\theta = T_\varphi^\varphi = (1 / \lambda r^4) \qquad ........(12)$$

The field equations in normal gauge for Lyra geometry for the metric (6) reduce to

$$- e^{-\beta} [\frac{1}{2} \mu^{11} + \frac{1}{4} (\mu^1)^2 - \frac{1}{4} \mu^1 \beta^1 - (\beta^1 / r) + (\mu^1 / r) + (1 / r^2)] + (1 / r^2) - \frac{3}{4} \beta_0^2 e^{-\gamma}$$

$$= (8\pi \eta^2 / r^2) - (8\pi / \lambda r^4) \qquad ........(13)$$

$$- e^{-\beta} [(\mu^1 / r) + \frac{1}{4} \mu^1 \gamma^1 + (\gamma^1 / r) + (1 / r^2)] + (1 / r^2) + \frac{3}{4} \beta_0^2 e^{-\gamma}$$

$$= (8\pi \eta^2 / r^2) - (8\pi / \lambda r^4) \qquad .....(14)$$

$$-e^{-\beta}[\tfrac{1}{2}\gamma^{11} + \tfrac{1}{4}(\gamma^{1})^{2} + (\mu^{1}/2r) - (\beta^{1}/2r) + (\gamma^{1}/2r) + \tfrac{1}{2}\mu^{11} + \tfrac{1}{4}(\mu^{1})^{2}$$
$$-\tfrac{1}{4}\mu^{1}\beta^{1} + \tfrac{1}{4}\mu^{1}\gamma^{1} - \tfrac{1}{4}\gamma^{1}\beta^{1}] + \tfrac{3}{4}\beta_{0}^{2}e^{-\gamma} = (8\pi/\lambda r^{4}) \quad \ldots(15)$$

$$-e^{-\beta}[\tfrac{1}{2}\gamma^{11} + \tfrac{1}{4}(\gamma^{1})^{2} - (\beta^{1}/r) + (\gamma^{1}/r) - \tfrac{1}{4}\gamma^{1}\beta^{1} + (1/r^{2})] + (1/r^{2}) + \tfrac{3}{4}\beta_{0}^{2}e^{-\gamma}$$
$$= (8\pi\eta^{2}/r^{2}) - (8\pi/\lambda r^{4}) \quad \ldots(16)$$

### 3. Solutions in the weak field approximations:

At this stage, let us consider the weak field approximations and assume that

$$e^{\gamma} = 1 + f(r), \quad e^{\beta} = 1 + g(r), \quad e^{\mu} = 1 + h(r) \quad \ldots(17)$$

Here the functions f, g and h should be computed to first order in $\eta^{2}$, $(8\pi/\lambda)$ and $\beta_{0}^{2}$.

In these approximations eq. (13) - (16) take the following forms as

$$-\tfrac{1}{2}h^{11} + (g^{1}/r) - (h^{1}/r) + (g/r^{2}) - \tfrac{3}{4}\beta_{0}^{2} = (8\pi\eta^{2}/r^{2}) - (8\pi/\lambda r^{4}) \quad \ldots(18)$$

$$-(f^{1}/r) - (h^{1}/r) + (g/r^{2}) + \tfrac{3}{4}\beta_{0}^{2} = (8\pi\eta^{2}/r^{2}) - (8\pi/\lambda r^{4}) \quad \ldots(19)$$

$$-\tfrac{1}{2}h^{11} - \tfrac{1}{2}(h^{1}/r) - \tfrac{1}{2}f^{11} - \tfrac{1}{2}(f^{1}/r) + \tfrac{1}{2}(g^{1}/r) + \tfrac{3}{4}\beta_{0}^{2} = (8\pi/\lambda r^{4}) \quad \ldots(20)$$

$$-\tfrac{1}{2}f^{11} - (f^{1}/r) + (g^{1}/r) + (g/r^{2}) + \tfrac{3}{4}\beta_{0}^{2} = (8\pi\eta^{2}/r^{2}) - (8\pi/\lambda r^{4}) \quad \ldots(21)$$

Hence the solutions of f, g and h are

$$f = \tfrac{1}{2}\beta_{0}^{2}r^{2} - (8\pi/3\lambda r^{2}) - 8\pi\eta^{2} \quad \ldots(22)$$

$$g = \tfrac{1}{4}\beta_{0}^{2}r^{2} + (8\pi/3\lambda r^{2}) + 8\pi\eta^{2} \quad \ldots(23)$$

$$h = 8\pi\eta^{2} - (8\pi/3\lambda r^{2}) \quad \ldots(24)$$

Thus in the weak approximations, the higher dimensional monopole metric in Lyra geometry takes the following form

$$ds^{2} = [1 + \tfrac{1}{2}\beta_{0}^{2}r^{2} - (8\pi/3\lambda r^{2}) - 8\pi\eta^{2}]dt^{2} - [1 + \tfrac{1}{4}\beta_{0}^{2}r^{2} + (8\pi/3\lambda r^{2}) + 8\pi\eta^{2}]dr^{2}$$
$$- r^{2}\Omega_{2}^{2} - [1 + 8\pi\eta^{2} - (8\pi/3\lambda r^{2})]d\psi^{2}$$
$$\ldots(25)$$

## 4. Motion of a test particle:

Let us consider a relativistic particle of mass m moving in the gravitational field of the monopole described by eq. (25).
The Hamilton – Jacobi ( H-J) equation is [12]

$$(1/A)(\partial S/\partial t)^2 - (1/B)(\partial S/\partial r)^2 - (1/r^2)[(\partial S/\partial x_1)^2 + (\partial S/\partial x_2)^2] -$$

$$(1/C)(\partial S/\partial \psi)^2 + m^2 = 0 \qquad \ldots\ldots\ldots(26)$$

$$A = [1 + \tfrac{1}{2}\beta_0^2 r^2 - (8\pi/3\lambda r^2) - 8\pi \eta^2]$$

$$B = [1 + \tfrac{1}{4}\beta_0^2 r^2 + (8\pi/3\lambda r^2) + 8\pi \eta^2]$$

$$C = [1 + 8\pi \eta^2 - (8\pi/3\lambda r^2)]$$

and $x_1, x_2$ are the co ordinates on the surface of the 2 – sphere.

Take the ansatz $S(t, r, x_1, x_2, \psi) = -E.t + S_1(r) + p_1.x_1 + p_2.x_2 + J.\psi$ …..(27)

as the solution to the H-J eq. (26).

Here the constants E, J are identified as the energy and five dimensional velocity and $p_1, p_2$ are momentum of the particle along different axes on 2 – sphere
with $p = (p_1^2 + p_2^2)^{1/2}$, as the resulting momentum of the particle.
Now substituting (27) in (26), we get

$$S_1(r) = \varepsilon \int [B(E^2/A) - (p^2/r^2) - (J^2/C) + m^2]^{1/2} dr \quad (\text{where } \varepsilon = \pm 1) \quad \ldots.(28)$$

In H-J formalism, the path of the particle is characterized by [12]

$$(\partial S/\partial E) = \text{constant}, (\partial S/\partial p_i) = \text{constant } (i = 1,2), (\partial S/\partial J) = \text{constant} \quad \ldots\ldots(29)$$

Thus we get (taking the constants to be zero without any loss of generality),

$$t = \varepsilon \int (\sqrt{BE}/A)[(E^2/A) - (p^2/r^2) - (J^2/C) + m^2]^{-1/2} dr \qquad \ldots\ldots(30)$$

$$x_i = \varepsilon \int (\sqrt{B} p_i / r^2)[(E^2/A) - (p^2/r^2) - (J^2/C) + m^2]^{-1/2} dr \qquad \ldots..(31)$$

$$\psi = \varepsilon \int (\sqrt{B} J / C)[(E^2/A) - (p^2/r^2) - (J^2/C) + m^2]^{-1/2} dr \qquad \ldots\ldots.(32)$$

From (30), we get the radial velocity as

$$(dr/dt) = (A/\sqrt{BE})[(E^2/A) - (p^2/r^2) - (J^2/C) + m^2]^{1/2} \qquad \ldots\ldots(33)$$

Now the turning points of the trajectory are given by $(dr/dt) = 0$ and as a consequence the potential curves are

$$(E/m) = [A \{( p^2/m^2r^2) + (J^2/Cm^2 ) - 1 \}]^{½} \qquad \ldots\ldots.(34)$$

Thus the extremals of the potential curve are the solution of the equation:

$$[\beta_0^{2} (J^{2} - m^{2})] r^{6} + [\beta_0^{2} (p^2 - 1)] r^{4} + (32 \pi p^2 / 3\lambda )$$

$$= r^{2} [(2 p^{2}(1 - 8\pi\eta^{2}) - (32 \pi J^2 / 3\lambda )]$$

[ neglecting $\eta^{4}$, $\beta_0^{4}$, $(8\pi / \lambda)^2$ and their product terms ]

This equation has real solution provided

$$p^{2}(1 - 8\pi\eta^{2}) > (16 \pi J^2 / 3\lambda ) \qquad \ldots\ldots\ldots.(35)$$

So the bound orbits are possible for the test particle i.e. particle can be trapped by global monopole.

## 5. Summary:

In this paper we have obtained an approximate solutions around a global monopole resulting from breaking of a global S0(3) symmetry in a five dimensional space time based on Lyra geometry. We note that our higher dimensional static metric is not locally flat and hence it represents a monopole [13]. We have also studied the motion of a test particle in gravitational field of the higher dimensional global monopole based on Lyra geometry using Hamilton – Jacobi formalism. We have shown that our monopole exerts gravitational force, which is attractive in nature.